\documentclass[conference]{IEEEtran}
\usepackage{cite}
\usepackage{amsmath,amssymb,amsfonts}
\usepackage{algorithmic}
\usepackage{graphicx}
\usepackage{textcomp}
\usepackage{makecell}
\usepackage{listings}
\usepackage{hhline}
\usepackage{xcolor}
\usepackage[format=plain, belowskip=-7pt]{caption}
\usepackage{url}

\begin{document}
\title{Optimisation of an FPGA Credit Default Swap engine by embracing dataflow techniques}

\author{\IEEEauthorblockN{Nick Brown}
\IEEEauthorblockA{EPCC at the University of Edinburgh\\
The Bayes Centre, 47 Potterrow,\\
Edinburgh, UK\\
Email: n.brown@epcc.ed.ac.uk}
\and
\IEEEauthorblockN{Mark Klaisoongnoen}
\IEEEauthorblockA{EPCC at the University of Edinburgh\\
The Bayes Centre, 47 Potterrow,\\
Edinburgh, UK}
\and
\IEEEauthorblockN{Oliver Thomson Brown}
\IEEEauthorblockA{EPCC at the University of Edinburgh\\
The Bayes Centre, 47 Potterrow,\\
Edinburgh, UK}
}

\maketitle

\begin{abstract}
Quantitative finance is the use of mathematical models to analyse financial markets and securities. Typically requiring significant amounts of computation, an important question is the role that novel architectures can play in accelerating these models in the future on HPC machines. In this paper we explore the optimisation of an existing, open source, FPGA based Credit Default Swap (CDS) engine using High Level Synthesis (HLS). Developed by Xilinx, and part of their open source Vitis libraries, the implementation of this engine currently favours flexibility and ease of integration over performance.

We explore redesigning the engine to fully embrace the dataflow approach, ultimately resulting in an engine which is around eight times faster on an Alveo U280 FPGA than the original Xilinx library version. We then compare five of our engines on the U280 against a 24-core Xeon Platinum Cascade Lake CPU, outperforming the CPU by around 1.55 times, with the FPGA consuming 4.7 times less power and delivering around seven times the power efficiency of the CPU.

%We first describe optimisation strategies, with around a 50 times runtime difference between the first and final version of the engine on FPGAs. This is then followed by an analysis of the impact of using different numerical precision and techniques. Subsequently, performance and power efficiency of the engine on a Xilinx Alveo U280 are compared against a 24 core Xeon Platinum CPU, with the FPGA outperforming the CPU both in terms of raw performance and power efficiency. The result of this work is a comparison, set of techniques, and lessons learnt that both apply to this specific CDS engine on FPGAs, and are also of interest more widely in accelerating other quantitative finance and high performance codes on reconfigurable architectures.
\end{abstract}

\IEEEpeerreviewmaketitle
\section{Introduction}
% important to get good performance and explore lots of options quickly

Quantitative finance analyses financial markets and securities using mathematical models. Not only is the ability to stream in data and generate immediate decisions important, but so is the capability to perform batch processing of financial data on HPC machines, for instance overnight, which must still occur withing specific time constraints. Finance, especially high frequency trading \cite{leber2011high}, has long been associated with Field Programmable Gate Arrays (FPGAs) which are a form of reconfigurable architecture. Providing a large number of configurable logic blocks sitting within a sea of configurable interconnect, FPGAs are hugely versatile. Whilst the adoption of this technology in finance has enjoyed success, to date it has been specialist and requiring considerable expertise to deploy successfully. However, the significant investment made by vendors in recent years around the software ecosystem for programming FPGAs, along with the availability of more capable hardware, has the potential to drive increased ubiquity and lower the barrier to entry especially for HPC.

The borrowing of money is a very common financial activity, and the Credit Default Swap (CDS) mechanism enables investors to offset their credit risk with that of another investor. This can be thought of as an insurance policy for the non-payment of loans, where if a lender believes there to be a risk that a loan will be defaulted upon, then a CDS can be bought from another investor who will reimburse the original lender if such occurs. A premium payment is required to maintain the CDS contract, which is effectively profit for the CDS seller if the loan is repaid successfully. This is a very common financial mechanism, and risk must be weighed by both buyers and sellers to determine the value of a CDS.

There are a number of key mathematical formula \cite{hull2009options} that analysts use to estimate CDS risk and value. Based upon such mathematics Xilinx developed an open source CDS engine for FPGAs in their Vitis library \cite{vitis-libraries}. However, Xilinx have made design decisions which favour flexibility and simplicity of integration, which limits the overall performance. In this paper we explore optimisation of this implementation using dataflow algorithmic techniques which fully suit the FPGA. The paper is organised as follows, in Section \ref{sec:background} we describe the background to CDS in more detail, specifics of the kernel we focus on in this work, and the FPGA technologies used. This is then followed by Section \ref{sec:development} which explores the optimisation of the Xilinx CDS engine, and steps required to obtain good performance. Section \ref{sec:multi-kernel} then compares the performance and power efficiency of all cores of the CPU, against five of our optimised CDS engines on the FPGA, followed by Section \ref{sec:conclusions} where we draw conclusions and discuss further work.

\section{Background}
\label{sec:background}

The Vitis Platform \cite{vitis} is an FPGA programming eco-system developed by Xilinx which promises to deliver an environment that lowers the barrier to entry in programming FPGAs and their use in accelerating applications. A major part of this is High Level Synthesis (HLS), which enables the development of FPGA kernels in C or C++, and there are also numerous other components to Vitis including access to convenient emulation, profiling tooling, and open source libraries. Open source libraries are especially important for domains such as finance because, whilst finance has been a user of FPGAs for many years \cite{de2015fpga} \cite{ruan2013fpga}, there is limited sharing of code due to commercial sensitivities and traditionally high barriers to entry in developing tuned financial FPGA kernels.

\begin{figure*}[htbp]
\centering
\includegraphics[scale=0.50]{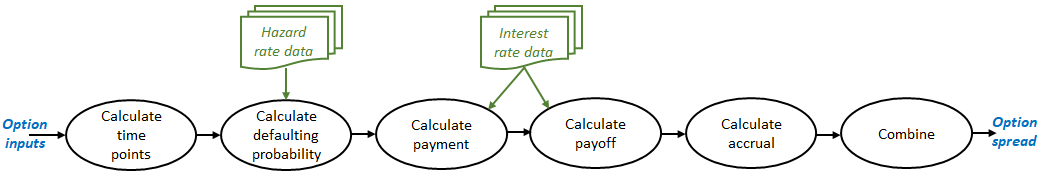}
\caption{Flowchart illustration of the structure of the Xilinx CDS FPGA engine}
\label{fig:xilinx_cds}
\end{figure*}

One area of focus for the Vitis open source libraries is that of quantitative finance and it contains, amongst other functionality, a CDS engine implementation. This is part of a wider commitment to financial applications, where more widely Xilinx have developed components including their Accelerated Algorithmic Trading (AAT) platform to aid in building high frequency trading solutions.

%with the API designed around three levels of abstraction, with subsequent levels designed to build upon support from previous ones. Aimed at providing a reusable API for programmers, Level 1 contains the low-level mathematical building blocks, level 2 are the applications which run on the FPGAs themselves, and level 3 are the hosts-side abstractions to launch and manage execution on the FPGA. Provided in level two of this Xilinx open source library is a CDS engine

% High frequency trading with Xilinx streamling library (get correct name!)

\subsection{Credit Default Swap (CDS) engine}
A Credit Default Swap (CDS) engine calculates the value known as \emph{spread}, which is the annual amount in basis points that the CDS protection buyer should pay the protection seller. Dividing this basis points number by 100 results in a percentage of the overall loan. There are three main inputs to the model, the interest and hazard rates which are constant for the model run, and a vector of options. The interest rate, or term structure, is expressed as a list of percentages of interest payable on the loan in a given time frame. The hazard rate expresses the likelihood that the loan will default by a specific point in time. Elements comprising these input values consist of two numbers, the point in time (fraction of a year), and the interest or hazard value itself.

% XIlix based on Quantlib QuantLib \cite{firth2004use}

Once this constant data has been loaded, then options are loaded into the engine. Each option comprises three elements of data, the maturity date (when the loan is expected to be repaid, effectively the end of the CDS), the frequency of payment, and the recovery rate (the percentage of the loan not repaid by the CDS). Based upon this data a spread result is calculated for each option, which the financial analysts then use to determine the price, or fee, of the CDS itself. 

A flow chart of how the existing open source Xilinx FPGA kernel operates is illustrated in Figure \ref{fig:xilinx_cds} where, for each option, the model first determines a set of distinct time points. These extend to the maturity date (the end of the CDS) and each subsequent component of the model loops over these time points. The first significant option calculation performed for each time point is the probability that the loan has defaulted by that point in time, which involves accumulating the hazard rate constant data up until this time. The next two calculations, the present value of expected payments on the loan and the present value of the expected payoff (the amount the seller must pay the buyer if the loan defaults at this point in time), requires the interest rate constant data. Lastly, the accrued protection at each specific point in time is calculated, which is the CDS insurance that has been paid for but not yet received (as premiums are paid ahead of time). Once the default probability, payment, payoff, and accrual terms are calculated, then the values are combined to calculate the overall spread, and this is returned from the model for the option in question. 

This procedure is then repeated, with the same interest and hazard rate constant data, for many different option configurations. The code structure is ideal for being expressed in dataflow style, however whilst the Xilinx implementation pipelines the individual loops it does not dataflow these, and as such the components making up the overall flowchart of Figure \ref{fig:xilinx_cds} run sequentially. All calculations involve double precision floating point.

\subsection{Experimental setup}

The experiments conducted in this paper use a Xilinx Alveo U280 card, which contains an FPGA with 1.3 million LUTs, 4.5MB of BRAM, 30MB of UltraRAM (URAM), and 9024 DSP slices. This PCIe card also contains 8GB of High Bandwidth Memory (HBM2) and 32GB of DRAM on the board. We use Vitis version 2020.2, and all code is compiled at optimisation level three for both the CPU and FPGA HLS code, and GCC version 7.4. For all experiments, 1024 interest and hazard rates are used, and results are averaged over three runs. The overhead of data transfer via PCIe is included for all FPGA results, which nevertheless represents a small part of the overall execution time. For performance comparison against the CPU, we use a 24-core Xeon Platinum (Cascade Lake) 8260M and implemented a bespoke version of the engine in C++ with OpenMP for multi-threading.

\section{HLS CDS engine optimisation}
\label{sec:development}

\begin{figure}[htbp]
\centering
\includegraphics[scale=0.50]{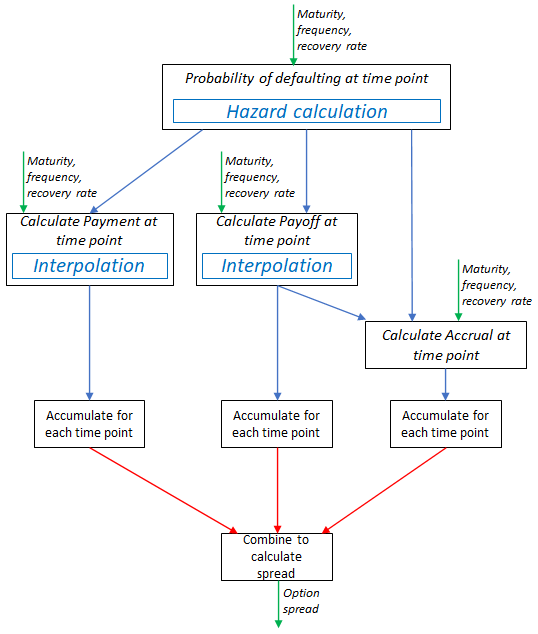}
\caption{Illustration of our CDS dataflow architecture}
\label{fig:dataflow_architecture}
\end{figure}

We developed a new version of the engine using an explicit dataflow style via the HLS \emph{DATAFLOW} pragma. In this approach distinct dataflow regions are declared as functions, operating concurrently and connected to other dataflow functions via HLS streams. Data is streaming between these concurrently running functions, with Figure \ref{fig:dataflow_architecture} illustrating the reorganisation of the code into this dataflow style. Each block box of Figure \ref{fig:dataflow_architecture} represents a function running concurrently, with red arrows indicating streams of values that occur once per option, and the blue arrows represent streams for each time point. The regions that calculate the defaulting probability, payment, payoff, accrual, and accumulation of values operate for each time point, with the final summed values for the option then streamed to the last stage which combines them and streams out the option's spread result. Inside some dataflow regions there are sub calculations, for instance the hazard calculation or interpolation sub-steps that operate for each time point. The green arrows in Figure \ref{fig:dataflow_architecture} represent input or output scalar values provided to, or returned from, the engine for each option. External data is located in the Alveo U280's HBM2 memory, and in accordance with best practice \cite{vitis-optimisingperf} external data accesses are packed into widths of 512 bits. 

%Inside our HLS engine code, to calculate both the expected payments and present value, an interpolation on the interest rate data is required. As per Xilinx's CDS engine we call into to the linear interpolation functionality, which is a level 1 (mathematical support) kernel of the Vitis library. 

The hazard calculation in the Xilinx implementation, required for calculating the probability of defaulting at a specific time point, was a bottleneck due to a spatial dependency. This calculation accumulates individual probability calculations based upon the hazard rate constant data. The accumulation, a double precision add, requires seven cycles to complete. Therefore the pipelined loop had an Initiation Interval (II), the number of cycles before the next value can start to be processed, of seven because such a number of cycles had to elapse before the next value could start to be processed due to this dependence on the previously accumulated value. Consequently, in the Xilinx open source implementation the hazard calculation was only generating a value for one of every seven cycles, impacting the performance of subsequent calculations which rely on this value. 

\begin{lstlisting}[frame=lines,caption={Sketch of the updated hazard calculation to avoid spatial dependency},label={lst:updated_hazard}, numbers=left]
double sumVal = 0.0, values[7];
for (unsigned int j = 0; j < 7; j++) values[j]=0.0;

for (unsigned int i = 0; i < length/7; i++) {
#pragma HLS PIPELINE II=7
  for (unsigned int j = 0; j < 7; j++) {
#pragma HLS UNROLL
    values[j] += .....
  }
}

for (unsigned int j = 0; j < 7; j++) {
#pragma HLS PIPELINE II=7
    sumVal+=values[j];
}
\end{lstlisting}

Listing \ref{lst:updated_hazard} is an illustration of how we addressed the spatial dependency in the hazard calculation code. By replicating the target accumulation variable into an array of seven elements (\emph{values} in Listing \ref{lst:updated_hazard}), and working cyclically in chunks of 7 cycles, we can on average process seven independent additions every seven cycles. The outer loop (line 4) has an II of 7, and the inner loop (line 6) is completely unrolled. Therefore, every seven cycles seven double precision additions will be completed independently. The loop between lines 12 and 15 sums up these temporary values, and whilst this suffers the same spatial dependencies, the impact is minimal as this final loop only operates on 7 elements rather than the entire length. For brevity, we omit the handling of an uneven division of length 7 in Listing \ref{lst:updated_hazard}, this is included in our engine code.

The performance metric we use in this paper is the number of options that can be processed per second, for all experiments 1024 interest and hazard rates are used and reported results include the overhead of data transfer. Performance of this optimised version on the Alveo U280 is illustrated by the row \emph{Optimised Dataflow CDS engine} in Table \ref{fig-kernel-optimisation}. For comparison, the performance of the CPU C++ engine on a single core of a Xeon Platinum (Cascade Lake) 8260M CPU is also included, as is the existing Xilinx Vitis library CDS implementation. It can be seen that our initial optimised engine was around twice as fast as the Xilinx open source implementation, although falling slightly short of CPU single-core performance.

\begin{table}[h]

%\hspace*{-0.3in}
%\begin{center}
 \centering
\begin{tabular}{ | c c | }
\hline
\textbf{Description} & \makecell{\textbf{Performance}\\\textbf{(Options/second)}} \\\hline
Xeon Platinum CPU core & 8738.92 \\ \hline
Xilinx Vitis library CDS engine & 3462.53 \\ \hline
Optimised Dataflow CDS engine & 7368.42 \\ \hline
Dataflow inter-options & 13298.70 \\ \hline
Vectorisation of dataflow engine & 27675.67 \\ \hline
\end{tabular}
%\end{center}
\caption{Performance of different versions of our FPGA CDS engine, against that of a Cascade Lake Xeon Platinum CPU single-core and Xilinx Vitis library implementation}
\label{fig-kernel-optimisation}
\end{table}

The Xilinx CDS engine processed one option at a time, where input values for an option are loaded, the calculations then undertaken for each time point, and then the spread returned. In our optimised dataflow approach this meant that the dataflow region shuts-down and restarts between options, and in addition to the performance overhead of starting and stopping the dataflow region, the pipelines were also continually filling and draining. Consequently, we modified the engine to run continually between options. This required changing the input and output option parameters to be streams, rather than individual scalar values, and also involved each dataflow stage being aware of the overall number of options. This significantly improved our performance by almost two times to 13298 options per second, \emph{Dataflow inter-options} in Table \ref{fig-kernel-optimisation}, and for the first time we were out performing the CPU core. 

\begin{figure}[htbp]
\centering
\includegraphics[scale=0.40]{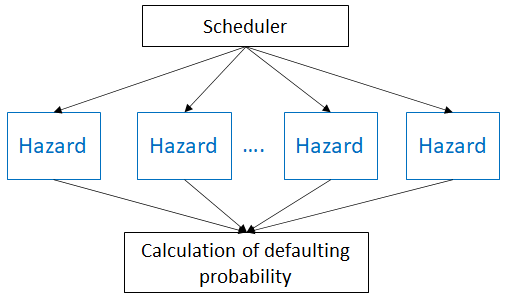}
\caption{Vectorisation of defaulting probability calculation}
\label{fig:vectorisation}
\end{figure}

The hazard calculation and linear interpolations of Figure \ref{fig:dataflow_architecture} involve nested loops. Therefore they require many cycles to produce a result for a single time point. Other dataflow stages, such as the accrual calculation or accumulations, do not share this feature and can generate a result per cycle, but as they depend upon data from such preceding stages, stalls frequently occurred. For that reason we replicated, or vectorised, those sub-functions which perform the hazard calculation or interpolation functionality, effectively running a number of them concurrently on different time points. 

Our approach is illustrated in Figure \ref{fig:vectorisation} for calculating the probability of loan defaults. Expressed as an HLS nested dataflow region, the \emph{scheduler} works round-robin style, streaming input data to the different functions cyclically, and the \emph{calculation of defaulting probability} then receives results cyclically and proceeds to process further. By working cyclically ordering of result consumption is maintained, and this approach improves the flow of data as the processing of multiple time points can run concurrently. The performance benefit of this optimisation is illustrated by the row \emph{Vectorisation of dataflow engine} in Table \ref{fig-kernel-optimisation}, where we replicated the hazard and interpolation calculations six times, which doubled performance. However this does increase resource usage, requiring additional logic for these replicated functions, and also additional dual-ported URAM storing the hazard and interest rate constant data. At this point, our FPGA engine was out-performing a CPU single-core by three times, and the Xilinx open source version by around eight times.

\section{Performance and power when scaling up}
\label{sec:multi-kernel}

We scaled up the number of CDS engines on the FPGA, being able to fit five onto the Alveo U280. There are no dependencies between calculations involving different options, and as such we decomposed based upon the options themselves, splitting the entire set up into \emph{N} chunks, where \emph{N} is the number of CDS engines. All engines require the full interest and hazard rate data, which is read in upon initialisation of the engine and stored in UltraRAM. Table \ref{fig-multi-kernel} illustrates the performance and power characteristics of our FPGA approach as we scale the number of engines. %It should be noted that FPGA performance includes the overhead of data transfer, and this is negligible overall.

We also compared against the full 24-core Xeon Platinum (Cascade Lake) 8260M CPU, using OpenMP to thread across the cores. This is reported by the first row of Table \ref{fig-multi-kernel} and it can be seen that in comparison to the single-core performance (see Table \ref{fig-kernel-optimisation}) the CPU code is scaling fairly poorly, where we have increased the core count by 24 times but the performance only increases by around nine times. It can be seen that, at five CDS engines, our FPGA approach is out performing all 24 cores of the Xeon Platinum CPU by around 1.55 times.

Table \ref{fig-multi-kernel} also illustrates that the FPGA running with five engines draws around 4.7 times less power than the CPU. Furthermore, it can be seen that the additional power overhead of adding extra FPGA engines is fairly minimal. Therefore the power efficiency is significantly higher on the FPGA, around seven times, than the CPU, as a larger number of FPGA kernels improve performance at little power cost.

\begin{table}[h]

%\hspace*{-0.3in}
%\begin{center}
 \centering
\begin{tabular}{ | c c c c | }
\hline
\textbf{Description} & \makecell{\textbf{Performance}\\\textbf{(Options/second)}} & \makecell{\textbf{Power draw}\\\textbf{(Watts)}} & \makecell{\textbf{Power efficiency}\\\textbf{(Options/Watt)}}\\\hline
24 core Xeon CPU & 75823.77 & 175.39 & 432.31 \\ \hline
1 FPGA engine & 27675.67 & 35.86 & 771.77 \\ \hline
2 FPGA engines & 53763.86 & 35.79 & 1502.20 \\ \hline
5 FPGA engines & 114115.92 & 37.38 & 3052.86 \\ \hline
\end{tabular}
%\end{center}
\caption{Performance and power when scaling the FPGA CDS engines on an Alveo U280, against 24-core Xeon CPU}
\label{fig-multi-kernel}
\end{table}

%When scaling from one engine to two the performance only increases by around 1.5 times, however when from going from two to four the performance increase is much closer to two. The reason is that, beyond one engine Vitis dynamically down-clocks the HLS IP blocks from 300 MHz (the default) to 266 Mhz in order to ensure that the design meets timing.

\section{Conclusions and further work}
\label{sec:conclusions}

In this paper we have explored the optimisation of the Xilinx open source Credit Default Swap (CDS) engine. Calculating the fee, or spread, of a series of options, such mathematical models are an important tool for financial analysts, and often placed under severe time constraints when generating results, hence requiring HPC. We described appropriate dataflow optimisations to the code and it is stark that, by moving to a fully dataflow style of algorithm, we were able to improve FPGA single kernel performance by around eight times. 

We compared the performance and power efficiency of up to five of our engines on an Alveo U280, against that of a 24-core Intel Xeon Platinum Cascade Lake CPU, and demonstrated that the FPGA delivers around 1.55 times greater performance at significantly reduced power consumption. This resulted in an overall power efficiency for the FPGA of seven times that of the CPU, and demonstrates the benefit of the FPGA for this type of workload.

Going forwards, further exploration around reduced precision, especially within the context of the future Xilinx Versal ACAP with AI engines for accelerating single precision floating point and fixed-point arithmetic, would be very interesting. Furthermore, combining our optimised CDS engine with Xilinx's high frequency trading AAT platform would be a useful next step to encourage more widespread adoption.

\section*{Acknowledgment}
The authors thank the ExCALIBUR H\&ES FPGA testbed for access to compute resource used in this work.

\bibliographystyle{IEEEtran}
\bibliography{references}

%\begin{thebibliography}{1}

%bibitem{IEEEhowto:kopka}

%\end{thebibliography}
\end{document}